\documentclass[sigconf]{acmart}

\usepackage[ruled,vlined]{algorithm2e}
\usepackage{multirow}
\usepackage{hhline}
\usepackage{graphicx}
\usepackage{pifont}
\usepackage{verbatim}
\usepackage{booktabs}
\usepackage{makecell}
\usepackage{hyperref}
\usepackage{siunitx}
\DeclareSIUnit[quantity-product = ]\percent{\char`\%}

\AtBeginDocument{%
  \providecommand\BibTeX{{%
    \normalfont B\kern-0.5em{\scshape i\kern-0.25em b}\kern-0.8em\TeX}}}

\acmConference[SIGGRPAH ASIA'23]{SIGGRPAH ASIA'23}{December 12--15, 2023}{Sydney}

\copyrightyear{2023}
\acmYear{2023}
\setcopyright{acmlicensed}\acmConference[SA Conference Papers '23]{SIGGRAPH Asia 2023 Conference Papers}{December 12--15, 2023}{Sydney, NSW, Australia}
\acmBooktitle{SIGGRAPH Asia 2023 Conference Papers (SA Conference Papers '23), December 12--15, 2023, Sydney, NSW, Australia}
\acmPrice{15.00}
\acmDOI{10.1145/3610548.3618148}
\acmISBN{979-8-4007-0315-7/23/12}

\definecolor{orchid}{rgb}{0.85, 0.44, 0.84}
\newcommand{\CHANGE}[1]{{{#1}}}

\newcommand{\HW}[1]{{#1}}

\begin{document}

\title{A Locality-based Neural Solver for Optical Motion Capture}

\fancyhead[ER]{\sffamily \footnotesize Xiaoyu Pan, Bowen Zheng, Xinwei Jiang, Guanglong Xu, Xianli Gu, \\ Jingxiang
Li, Qilong Kou, He Wang, Tianjia Shao, Kun Zhou and Xiaogang Jin}

\author{Xiaoyu Pan}
\email{panxiaoyu6@gmail.com}

\author{Bowen Zheng}
\email{zhengbowen.crist@gmail.com}

\affiliation{%
  \institution{State Key Lab of CAD \& CG, Zhejiang University; ZJU-Tencent Game and Intelligent Graphics Innovation Technology Joint Lab}
  \city{Hangzhou}
  \country{China}}

\author{Xinwei Jiang}
\email{wesleyjiang@tencent.com}

\author{Guanglong Xu}
\email{jasonglxu@tencent.com}

\author{Xianli Gu}
\email{shellygu@tencent.com}

\author{Jingxiang Li}
\email{jingxiangli@tencent.com}

\affiliation{%
  \institution{Tencent Games Digital Content Technology Center}
  \city{Shanghai}
  \country{China}}

\author{Qilong Kou}
\email{rambokou@tencent.com}

\affiliation{%
  \institution{ Tencent Technology (Shenzhen) Co., LTD}
  \city{Shenzhen}
  \country{China}}

\author{He Wang}
\email{he_wang@ucl.ac.uk}

\affiliation{%
  \institution{University College London (UCL)}
  \city{London}
  \country{United Kingdom}}

\author{Tianjia Shao}
\email{tianjiashao@gmail.com}

\author{Kun Zhou}
\email{kunzhou@acm.org}

\affiliation{%
  \institution{State Key Lab of CAD \& CG, Zhejiang University}
  \city{Hangzhou}
  \country{China}}

\author{Xiaogang Jin}
\authornote{Corresponding author.}
\affiliation{%
  \institution{State Key Lab of CAD \& CG, Zhejiang University; ZJU-Tencent Game and Intelligent Graphics Innovation Technology Joint Lab}
  \city{Hangzhou}
  \country{China}}
\email{jin@cad.zju.edu.cn}

\begin{abstract}

We present a novel locality-based learning method for cleaning and solving optical motion capture data. Given noisy marker data, we propose a new heterogeneous graph neural network which treats markers and joints as different types of nodes, and uses graph convolution operations to extract the local features of markers and joints and transform them to clean motions. To deal with anomaly markers (e.g. occluded or with big tracking errors), the key insight is that a marker's motion shows strong correlations with the motions of its immediate neighboring markers but less so with other markers, a.k.a. \textit{locality}, which enables us to efficiently fill missing markers (e.g. due to occlusion). Additionally, we also identify marker outliers due to tracking errors by investigating their acceleration profiles. Finally, we propose a training regime based on representation learning and data augmentation, by training the model on data with masking. The masking schemes aim to mimic the occluded and noisy markers often observed in the real data. Finally, we show that our method achieves high accuracy on multiple metrics across various datasets. Extensive comparison shows our method outperforms state-of-the-art methods in terms of prediction accuracy of occluded marker position error by approximately 20\%, which leads to a further error reduction on the reconstructed joint rotations and positions by 30\%. The code and data for this paper are available at \textcolor{magenta}{\textit{\url{https://github.com/non-void/LocalMoCap}}}.

\end{abstract}



\begin{teaserfigure}
\centering
    \includegraphics[width=\textwidth]{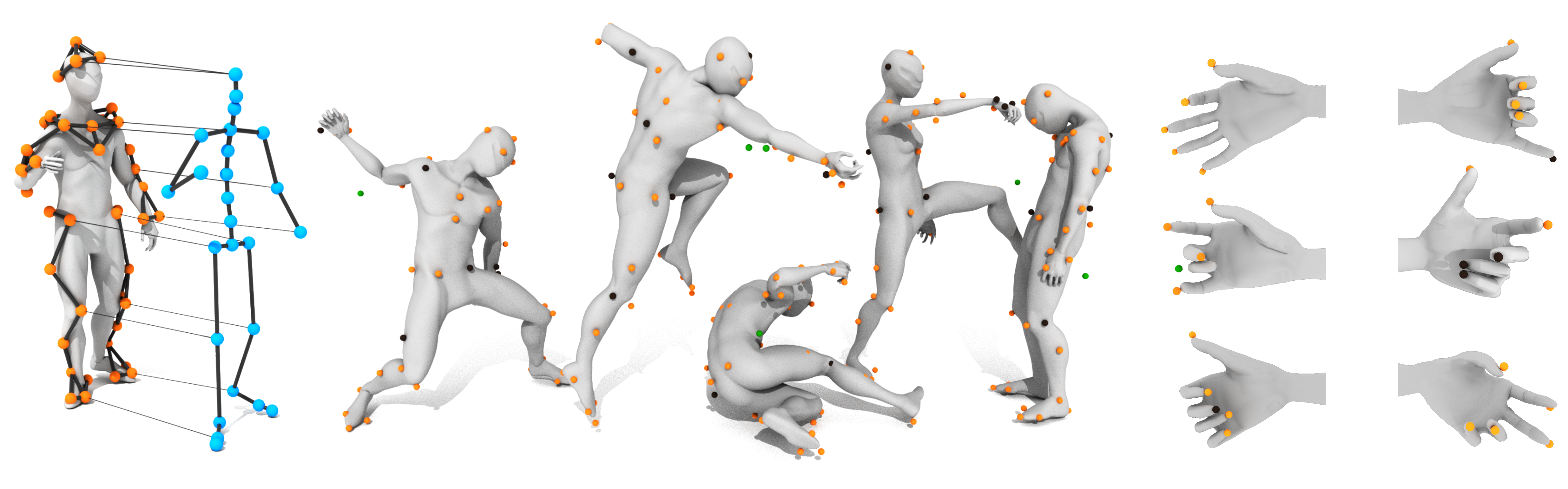}
    \caption{Given raw MoCap data, our method constructs a heterogeneous graph neural network (left) to solve large body motions (middle) and fine multi-fingered hand motions (right). Our method is robust to occlusions (black balls) and outliers (green balls).}
    \label{fig:teaser}
\end{teaserfigure}

\maketitle

\section{Introduction}
Motion capture (MoCap) is a popular industry solution for giving virtual characters realistic motions. It captures the motions of real-world actors using a variety of sensors. Because of its precision and flexibility, optical motion capture is still the industry standard for film and video game production to capture body and hand motions with highly coordinated movement. Even with high-fidelity and expensive motion capture equipment, raw data is inevitably contaminated by occlusions, position errors, and mislabelling. Despite the development of tools for cleaning and solving MoCap data, greater accuracy still requires manual fixing and adjustment. When it comes to complex motions like hand motions, even experienced animators find this process time-consuming.

Numerous efforts in the past have been made to automatically clean MoCap data and solve for motions. Early methods~\cite{herda2000skeleton,zordan2003mapping,kirk2004skeletal,aristidou2013real, li2009,BoLeRo,burke2016estimating,feng2014exploiting,liu2014automatic} heavily rely on rules abstracted from empirical observations and hand-crafted features. These methods can produce satisfactory results with specific patterns and noises via careful hand-tuning, but they constantly suffer from poor generalization onto real-world data, which contains complex noises, e.g. simultaneous occlusions of multiple markers, and varying duration of occlusions. Recently, data-driven methods~\cite{wang2016,xiao2015sparse,tits2018robust,holden18,soma,MoCapSolver,perepichka2019robust,pavllo2018real} have been employed to address the aforementioned limitations. For instance, MoCap-Solver~\cite{MoCapSolver} solves motions and reconstructs clean markers by separately encoding motion, marker configuration, and skeleton using different neural networks, and achieves the state-of-the-art results. However, dealing with noisy marker data with outliers and large simultaneous occlusions remains difficult. This is especially true for hand markers, which often have subtle motions involving frequent marker occlusions. 

Overall, we observe that the state-of-the-art data-driven methods~\cite{holden18,MoCapSolver} have three main limitations. First, they solve for motion and reconstruct clean markers using skinning functions, which ignore the complexity of marker motions (e.g., sliding across the body surface) and may introduce additional errors due to motion solving errors. Second, they often ignore the fine-grained correlations between markers, e.g. using one single fully connected network structure to encoder all markers in the same way, resulting in inaccurate solving results for certain large body motions and fine multi-fingered hand movements (see Figs. \ref{additional_results} and \ref{additional_results2}). Third, all methods have to assume/model data noise, where a common strategy is to use random sampling per frame. However, modeling noise data with per-frame random sampling does not account for long gaps caused by complex and occlusion-intensive movements and varying occlusion probability of different markers, potentially lowering solve accuracy on real data.

Three major challenges must be overcome in order to accurately solve for motions. (1) The ability to deal with complex occlusions that include simultaneous occlusions and occlusions of varying duration. Although neural-based methods can be used to learn from real-world data, using networks alone makes it difficult to learn complex patterns of marker motions, limiting its use for efficiently filling occluded marker positions. (2) Modeling the relationships between markers in a neural network. A critical task in quantifying such relationships is to establish links between markers with strong correlations while avoiding links to other markers. \CHANGE{Networks with fully connected structures do not explicitly model such relationships as they encode all markers in the same way.} (3) Creating realistic large-scale synthetic MoCap data with long occlusion gaps. Due to the high cost of MoCap systems, obtaining real noise data is difficult, necessitating data augmentation to generate synthetic noises. Actual MoCap sequences frequently have occlusion gaps ranging from 40 to 100 frames. The per-frame random sample method fails to generate occlusion gaps longer than eight frames (see Fig.~\ref{gap_distribution}), resulting in a mismatch between the synthetic and real-data distributions.

We present a data-driven method for cleaning optical MoCap data and solving for body and hand motions. A key aspect of our approach is leveraging marker locality by identifying neighbor markers with stable distances between them across the entire motion. Using this locality as a prior, we can obtain an initial estimate of occluded markers' positions using the distance matrix between neighbor markers via Euclidean distance matrix optimization, which reduces the difficulty of network learning intricate marker motions and significantly improves accuracy. To quantify the relationships between markers, we connect markers and related joints and construct edges between neighbor markers and parent joints to form a heterogeneous graph. We perform convolution operations to extract local features of markers and joints, which enables the network to accurately solve motions. In addition, we augment the dataset with data distributions similar to real data by sampling occlusion gaps of varying lengths for different markers based on actual occlusion probabilities and occlusion gap distributions. In summary, our paper makes the following three contributions:

\begin{itemize}
    \item A robust MoCap data occlusion filling method that integrates locality and learned priors to handle simultaneous occlusions and occlusions of varying durations.
    \item A heterogeneous graph neural network that solves motions by explicitly modeling the relationships between neighbor markers, allowing it to solve accurate large body motions as well as fine multi-fingered hand motions.
    \item A novel method for augmenting motion capture datasets that takes into account the statistical features of marker occlusion and generates data with a distribution similar to the real data.
\end{itemize}


\section{Related Work}

\CHANGE{Human pose estimation is an important and ongoing research topic in computer vision and computer graphics. While marker-free, image-based solutions have yielded promising results, marker-based motion capture remains popular due to its accuracy and flexibility~\cite{GRAB}. MoCap systems' high-fidelity body motion data~\cite{sigal2010humaneva,AMASS,CMU,trumble2017total} is useful in a variety of applications~\cite{survey1}, including action recognition~\cite{hua2023part,yan2018spatial}, action prediction~\cite{cao2020long,cui21}, motion synthesis~\cite{tevet2022human,chen2023executing} and image-based pose estimation~\cite{varol17_surreal,munea2020progress}.  However, raw MoCap data is inevitably contaminated with errors.
}

To solve motions from imperfect MoCap data, various methods for cleaning noises and solving motions have been proposed, which can be broadly classified into hand-crafted prior-based methods and data-driven models. The former are based on hard-coded empirical rules such as spatiotemporal continuity and bone-length consistency, which are implemented using skeleton templates~\cite{herda2000skeleton,zordan2003mapping,kirk2004skeletal}, Kalman filters~\cite{aristidou2013real, li2009,BoLeRo,burke2016estimating} and low-rank matrix~completion~\cite{feng2014exploiting,liu2014automatic}. However, the priors used in these methods make assumptions about the data and noise distributions, limiting their ability to handle real-world data with more complex noises.

Data-driven methods learn from a large database to acquire intrinsic knowledge of MoCap data, such as KD-tree~\cite{baumann11,tautges11}, local PCA~\cite{chai2005performance,liu2006estimation}, self-similarity~\cite{aristidou2018self}, sparse encoding \cite{wang2016,xiao2015sparse}, and model averaging~\cite{tits2018robust}. With the advancement of deep-learning, a number of neural-based methods have emerged~\cite{holden18,soma,MoCapSolver,perepichka2019robust,pavllo2018real}. \CHANGE{SOMA \cite{soma} uses a transformer-based network to automatically label the marker point cloud. While SOMA assigns unlabeled markers to specific body parts (such as the inner left wrist and right elbow), our method solves motions with labeled markers.
} The majority of works focus on repairing occluded markers and solving motions. \cite{pavllo2018real} uses auto-encoder-based models to recover occluded hand markers and solve hand motions. \cite{perepichka2019robust} presents a missing marker completion method by comparing motions generated by commercial software with those generated by a neural solver, the accuracy of these methods is determined by the neural solver. DeepMerf~\cite{madadi2021deep} fixes occluded markers with a denoising autoencoder and employs a cascading network to regress SMPL body parameters~\cite{SMPL-X} from joint positions estimated with an attention model. However, this method is based on the SMPL model and is difficult to apply to characters with different skeleton topologies. Holden~\cite{holden18} solves skeletal motions from MoCap data for characters with arbitrary skeletons using a simple forward neural network with residual blocks. Their method solves the motion and skeleton frame by frame and necessitates smoothing as a post-processing step for temporal motion continuity. MoCap-Solver~\cite{MoCapSolver} solves motions in a temporal window by encoding body shapes, marker distributions, and motions separately to ensure temporal continuity. Their method reconstructs clean markers based on skinning functions with solved motions, which ignores the complexity of marker motions (e.g. sliding over the body surface), and may induce additional errors due to motion solving errors. In contrast, our method fills the occlusion by incorporating hand-crafted priors and learned intrinsic priors, which accurately reconstruct occluded markers. Furthermore, previous methods~\cite{holden18,MoCapSolver} do not explicitly quantify the local feature of markers, which negatively affects solving result of certain large body motions and fine multi-fingered hand motions. Unlike them, our method extracts local features by constructing a heterogeneous graph that distinguishes markers and joints as different types of nodes and performing graph convolution operations on it, which significantly improves solving accuracy.

\begin{figure*}[htbp]
    \centering
    \includegraphics[width=\linewidth]{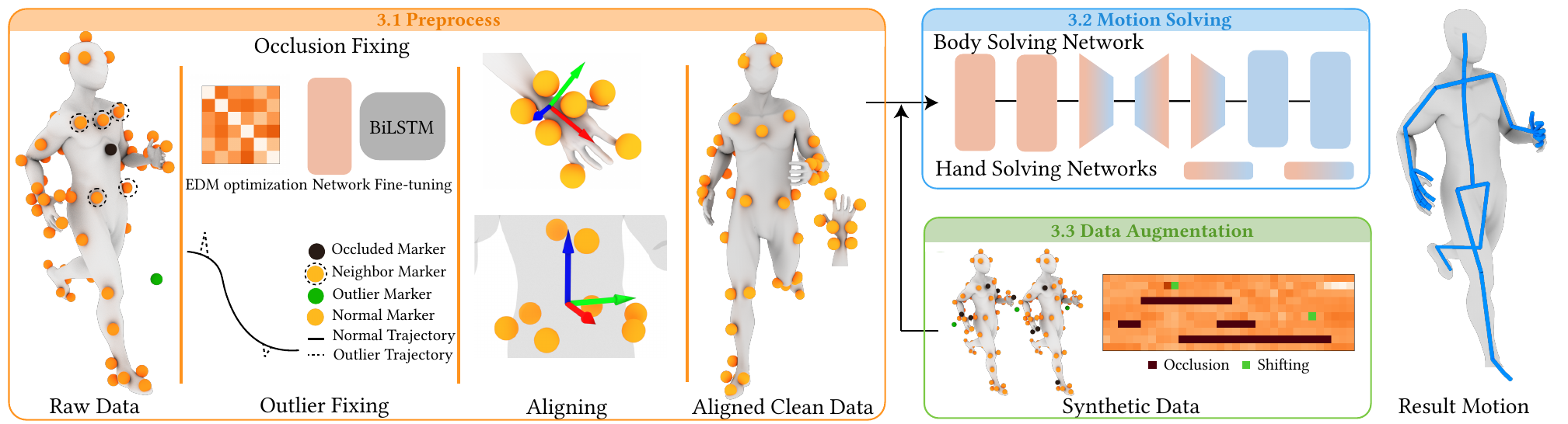}
    \caption{The pipeline of our method consists of three main modules. Left: Given the raw MoCap data, we clean the data by fixing occluded markers with the EDM algorithm with a bi-directional neural network and identifying outliers by locating the peaks on the marker acceleration curve, and remove the global transformation of body and hand markers by explicitly calculating their local coordinates using waist and wrist markers. Top right: Using the aligned clean body markers, we solve the body and hand motions separately using heterogeneous graph neural networks, whose details are shown in Fig.~\ref{network}. Bottom right: We augment our dataset by randomly sampling gaps and adding shifting based on actual MoCap data.}
    \label{pipeline}
\end{figure*}

\section{Method}

\label{method}

Given a sequence of raw body and hand marker data $M_{raw} \in \mathbb{R}^{T \times |M| \times 3}$, i.e., the positions of markers $M$ in a temporal window of $T$ frames, and their occlusion status $O \in [0,1]^{T \times |M|}$, which is a binary mask indicating the marker visibility, our method aims to solve the body and hand motions $Y \in \mathbb{R}^{|J| \times 9 + 3}$, which consists of the body's global translation and the rotation of every joint $J$, and the underlying skeleton $S \in \mathbb{R}^{|J| \times 3}$. In this section, we will first explain the neighbor markers and then present our pre-process method and the motion solving network. At the end, we describe the data augmentation algorithm.

Our method leverages spatial locality based on neighbor markers $\mathcal{N}(M)$. To find neighbor markers $\mathcal{N}(M_i)$ for marker $M_i$, we first compute the variances of pairwise distances between markers across the motions and choose $K$ markers with the lowest variances. The stable distances between $M_i$ and its neighbor markers help to efficiently solve for occluded marker positions. Furthermore, the neighbor markers can better represent motions of the corresponding body parts, assisting the motion solving networks in accurately solving body and hand motions.

Initially, the raw data $M_{raw}$ is subjected to a cleaning process that includes occlusion gap filling and outlier removal, resulting in refined data $M_{clean}$. This step contributes to enhancing accuracy of the motion solving network. The occlusion gap filling process is a two-step process in which we first calculate an approximate value $X_{EDM}$ of occluded markers based on their distances to neighbor markers, then fine-tune it using a bidirectional Long-Short Time Memory (biLSTM) network. Outliers are detected by identifying abnormal values in markers' acceleration curves and replaced by simple spline interpolation. To reduce training difficulty, we split body and hand markers and remove their global transformations by explicitly calculating local coordinate systems of wrist and waist markers. The resultant aligned clean markers are denoted by $M_{clean,local}$.

\begin{figure}[htbp]
    \centering
    \includegraphics[width=\linewidth]{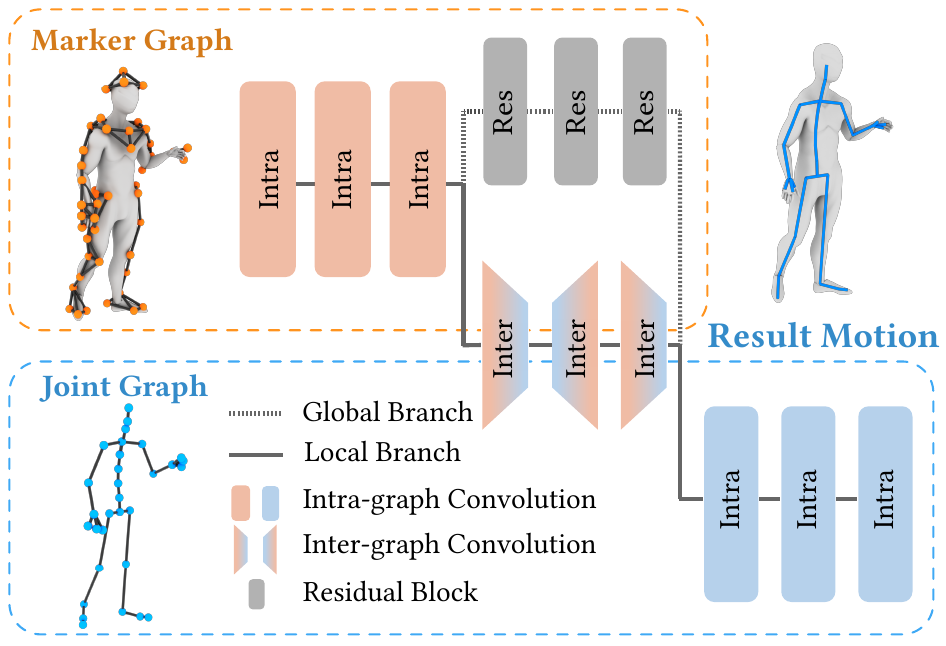}
    \caption{Our solving network structure. The network operates on a heterogeneous graph comprising the marker graph and joint graph. It first extracts the local features of markers, which are subsequently transferred to joint features and finally transformed into motion.}
    \label{network}
\end{figure}

 Next, we use heterogeneous graph neural networks to solve the motion frame-by-frame, as shown in Fig. \ref{network}. We train three networks with similar structures to separately solve body and hand motions. We begin by constructing a heterogeneous graph $\mathcal{G}$ comprised of the marker graph $\mathcal{G}_M$, whose nodes and edges represent marker and their spatial adjacency, and the joint graph $\mathcal{G}_J$, whose nodes and edges represent joints and bones, respectively. The connections between two graphs are based on spatial proximity of markers and joints. Our network employs both intra-graph and inter-graph convolution components, with the former operating within subgraphs and the latter transferring information between them. The network first extracts the markers' features from the marker graph $\mathcal{G}_M$ using a stack of intra-graph convolution layers, and then feeds these features into two branches: a global branch composed of residual blocks that extract global motion features, and a local branch composed of inter-graph convolution layers that extract local features of marker and joint nodes, representing the motion pattern of the body part. To obtain body motions, the features extracted by local and global branches are concatenated and fed into a stack of intra-graph convolution layers operating on joint graph $\mathcal{G}_J$.

Our approach generates synthetic data for data augmentation. We simulate occlusion gaps by randomly sampling gaps based on the occlusion probability of markers $\hat{p}_{occ,i}$ and the gap length distribution $\hat{g}_l$. Furthermore, we shift markers to simulate outliers based on given shifting probabilities and intensities $p_{shift}$ and $\sigma_{shift}$.

\subsection{Preprocess}

Real-world optical motion capture data is frequently plagued by occlusions and outliers, making it difficult for the solver to solve motions accurately. Furthermore, the global transformations of body and hand reduces the accuracy of the neural motion solver. To mitigate these issues, we preprocess the data by filling occlusions and remove outliers, and align markers to remove the global transformation, thereby facilitating more effective network training.

Marker occlusion is a common problem in optical motion capture systems caused by body occlusion or marker dropping. Occlusion gap filling is a highly non-linear problem due to the complexity of body motion and the uncertainty of the lengths and quantity of occlusion gaps. To address this issue, we propose a recurrent neural network-based approach. In contrast to previous learning-based methods, we use the Euclidean distance matrix optimization (EDM) algorithm to obtain an initial value for the network by solving the position of occluded markers based on the distance matrix of their neighbor markers. This approach is based on the observation that during motion, the distances between neighbor markers vary slightly. As a result, we calculate the position of the occluded marker based on its distances to neighbor markers in its visible frames.

Specifically, given a marker $M_i\in \mathbb{R}^3$ that is occluded between frames $s$ and $e$, we begin by identifying $K$ visible neighbor markers $M_j \in \mathcal{N}(M_i)$ between these frames. We then compute distance matrices $D_{i}^s, D_{i}^e \in \mathbb{R}^{(1+K)\times(1+K)}$ for $M_i$ and its neighbor markers at the start and end of the occlusion gap, respectively. We use linear interpolation to estimate the distance matrix at frame $t$ within this interval: $D_{i}^t=\frac{(e-t)D_{i}^s+(t-s)D_{i}^e}{e-s}$. \HW{Formally, $D_{i}^s$ and $D_{i}^e$ lie in a space of symmetric positive definite matrices where interpolation should follow geodesics. In practice, we found that linear interpolation provides a great approximation of the distance matrix at frame $t\in[s, e]$, as the distance between $M_i$ and neighbor markers vary only slightly. Furthermore, instead of treating neighboring markers as rigid bodies by using fixed distance matrices, the interpolated matrix reflects subtle change of distances between markers due to markers sliding over the skin surface or dynamic effects of soft body tissues.} Following that, we employ an EDM optimization method~\cite{zhou2020} to compute occluded marker's position  $M^{t}_{i,EDM}$. Formally, the EDM optimization algorithm is defined as follows:


\begin{equation}
    \begin{split}
        \min_{M^t_{EDM,i}} f(M_{EDM,i}^t,M_j^t)= \sum_{M_j\in \mathcal{N}(M_i)} {\big|} ||M_{EDM,i}^t-M_j^t||^2-D^t_i[i,j] {\big|} .
    \end{split}
\label{eq:EDM}
\end{equation}

To solve Eq.~\ref{eq:EDM}, the optimization is divided into two stages, where we first compute an initial guess on a set of marker positions $P_i^t\in \mathbb{R}^{(K+1)\times 3}$ that satisfies $D_i^t$, then align $P_i^t$ with $\mathcal{N}(M_i)$ by a rigid transformation. To calculate $P_i^t$, where one marker position corresponds to the occluded marker and the others correspond to its neighbor markers, we compute their initial positions using the Multi-Dimensional Scaling method~\cite{torgerson1952multidimensional} as follows: 
\begin{equation}
    \begin{split}
        P^t_i=[\mathbf{p}_1,...,\mathbf{p}_{r}]=diag(\sqrt{\lambda_1},...,\sqrt{\lambda_{r}})[\mathbf{x}_1,...,\mathbf{x}_{r}]^T,
    \end{split}
\end{equation}
where $\mathbf{p}_i$ represents the points in $P_i^t$, the eigenvalues $\lambda_i$ and corresponding eigenvectors $\mathbf{x}_i$ are from eigen decomposition of the MDS embedding $-\frac{1}{2}(JD_t^iJ)$, where $J=I-\frac{1}{K+1}\mathbf{1}\mathbf{1}^T$ is the centering matrix with $I$ being the identity matrix and $\mathbf{1}$ being vector of all ones and $r$ is the rank of $JDJ$. 

Next, we align the initial guess $P_i^t$ to $\mathcal{N}(M_i)$ using rigid transformation by solving the well-known Rotation Orthogonal Procrusts Problem~\cite{zhang2010rigid}. For simplicity, we use a closed-form solution by Singular Value Decomposition (SVD).

However, the distances between markers do not fully represent the complex motion patterns of markers. Next, we fine-tune the EDM result using a recurrent neural network that learns from a large dataset. The network is made up of an intra-convolution layer that extracts markers' local features and a biLSTM layer that uses temporal information. In the following section, we will go over the intra-convolution layer in more detail. The network takes as input the aligned marker positions and their occlusion status, denoted as $[M_{EDM,local,i}^t,0]$ for occluded markers and $[M_{clean,local,i}^t,1]$ for visible markers. The network estimates the offset$M^t_{off,i}$ from the EDM results to the ground truth using the following loss function:
\begin{equation}
    \begin{split}
        \mathcal{L}_{occ}=\sum ||(1-O^t_i) \odot 
        (\hat{M}^t_{clean,local,i}-(M^t_{EDM,local,i}+M^t_{off,i}))||,
    \end{split}
\end{equation}
 where $\hat{M}_{clean,local,i}^t$ is ground truth, and $\odot$ is element-wise product.

Another critical issue that can affect the quality of optical MoCap data is outlier. A common method for reducing such jittery movements in data is to apply a filter to the entire sequence, such as the Savitzsky-Golay filter~\cite{savitzky1964} used in~\cite{holden18}, which may reduce the fidelity of outlier-free parts. To reduce the affection to these parts, we identify the outliers and only fix the frames near where the outlier appears. In contrast to normal markers, which have smoother movement patterns, outliers have sudden jumps or accelerations due to labeling errors or other sources of noise. As a result, we use movement smoothness as a criterion to identify outliers, which are defined as markers with acceleration peaks that exceed a certain threshold. We remove the marker sequence around the outlier and fill the gap with spline interpolation to improve motion fidelity. 

To remove global transformation, we use a straightforward yet effective method for aligning body and hand markers to the local space. Similar with~\cite{MoCapSolver}, we select markers around rigid body parts, such as the waist and wrist. However, we observe that even these markers are not rigid bodies and exhibit slight non-rigid deformations, potentially causing instability in the rigid body alignment. To tackle this issue, we explicitly compute local coordinate systems using relationships between reference markers. A more detailed explanation can be found in the supplementary material.

\subsection{Motion Solving}

We construct a heterogeneous graph $\mathcal{G}=(\mathcal{G}_M,\mathcal{G}_J,\mathcal{A}_{MJ})$, using the markers $M$ and joints $J$. This graph comprises the marker graph $\mathcal{G}_M$, joint graph $\mathcal{G}_J$, and the adjacency matrix $\mathcal{A}_{MJ}$ representing the edges between the two subgraphs. The marker graph is denoted as $\mathcal{G}_M=(\mathcal{V}_M, \mathcal{E}_M)$, where $\mathcal{V}_M$ is the set of graph nodes indicating markers, and $\mathcal{E}_M$ denotes its edges connecting neighbor markers \CHANGE{(see the second paragraph of Section \ref{method})}. In the joint graph $\mathcal{G}_J=(\mathcal{V}_J, \mathcal{E}_J)$, each node in $\mathcal{V}_J$ represents a joint and edges in $\mathcal{E}_J$ are identical to the connections between joints. \CHANGE{The connections are dictated by the target skeleton structure.} We construct the adjacency matrix $\mathcal{A}_{MJ}$ based on the distance between joints and markers at T-pose, where $\mathcal{V}_{M,i}$ and $\mathcal{V}_{J,i}$ are connected if their corresponding marker and joints' distance is below a threshold. To represent the global transformation, we add an additional node to $\mathcal{G}_J$ that connects to every other nodes $\mathcal{V}$ in both $\mathcal{G}_J$ and $\mathcal{G}_M$.

The inputs of our network are positions of aligned markers and their visibility at frame $t$: $[M_{clean,local}^t,O^t]$, which are mapped to their corresponding nodes $\mathcal{V}_M$ in $\mathcal{G}_M$. The output of the network comprises features of joint nodes $\mathcal{V}_J$ in $\mathcal{G}_J$. Specifically, the output of $\mathcal{V}_{J,i}$ contains the rotation matrix $Y^t_{i}\in \mathbb{R}^9$ and the joint offset $S^t_{i}\in \mathbb{R}^3$. Since the network does not guarantee orthogonality, we use Gram-Schmidt orthogonalization as a post-process for the rotation matrix. The first three dimensions of the translation node represent the global translation. To ensure bone length consistency, the final joint offset is computed as the average of estimated joint offsets across the motions, denoted by $S_{i}=mean(S^1_{i},...,S^{T}_{i})$.

Our convolution operations are built based on the skeletal convolution in~\cite{aberman2020skeleton}. \CHANGE{To gather information from neighboring nodes, the operation employs a learned filter. There are some distinctions between intra- and inter-convolutions: the former aggregates the local features of the nodes' neighbors of the same type, whereas the latter aggregates the local features of their neighbors of a different type.  A marker node in our scenario gathers information from neighboring marker nodes in intra-convolution and aggregates features of linked joint nodes in inter-convolution. 
} The convolution operation is illustrated below:
\begin{equation}
    \begin{split}
        f'_{\mathcal{V}_i}=\sum_{\mathcal{V}_j \in \mathcal{N}(\mathcal{V}_i)}W_{ij}f_{\mathcal{V}_j} + b_i,
    \end{split}
\end{equation}
where $f_{\mathcal{V}_i}$ and $f'_{\mathcal{V}_i}$ are the input and output features of node $\mathcal{V}_i$, respectively. $\mathcal{N}(\mathcal{V}_i)$ is the set of its neighbors, and $W_{ij}, b_i$ are the learned filters and biases, respectively. It should be noted that unlike traditional convolution operations, our convolution operation employs different filters and biases for different edges.

We train the network using the following loss with three terms:
\begin{equation}
    \begin{split}
        \mathcal{L}_{solving}=\lambda_{M}(Y-\hat{Y})+\lambda_{S}(S-\hat{S})+\lambda_{FK}(FK(Y,S)-FK(\hat{Y},\hat{S})),
    \end{split}
\end{equation}
where the first two terms push the estimated motion and skeleton $Y, S$ to their ground truths $\hat{Y}, \hat{S}$, and $FK()$ is the forward kinematic function computing global positions of joints according to the motion and skeleton, and $\lambda_{M}, \lambda_{S}, \lambda_{FK}$ are predefined weight factors.

\subsection{Data Augmentation}

\begin{figure}[htbp]
    \centering
    \includegraphics[width=\linewidth]{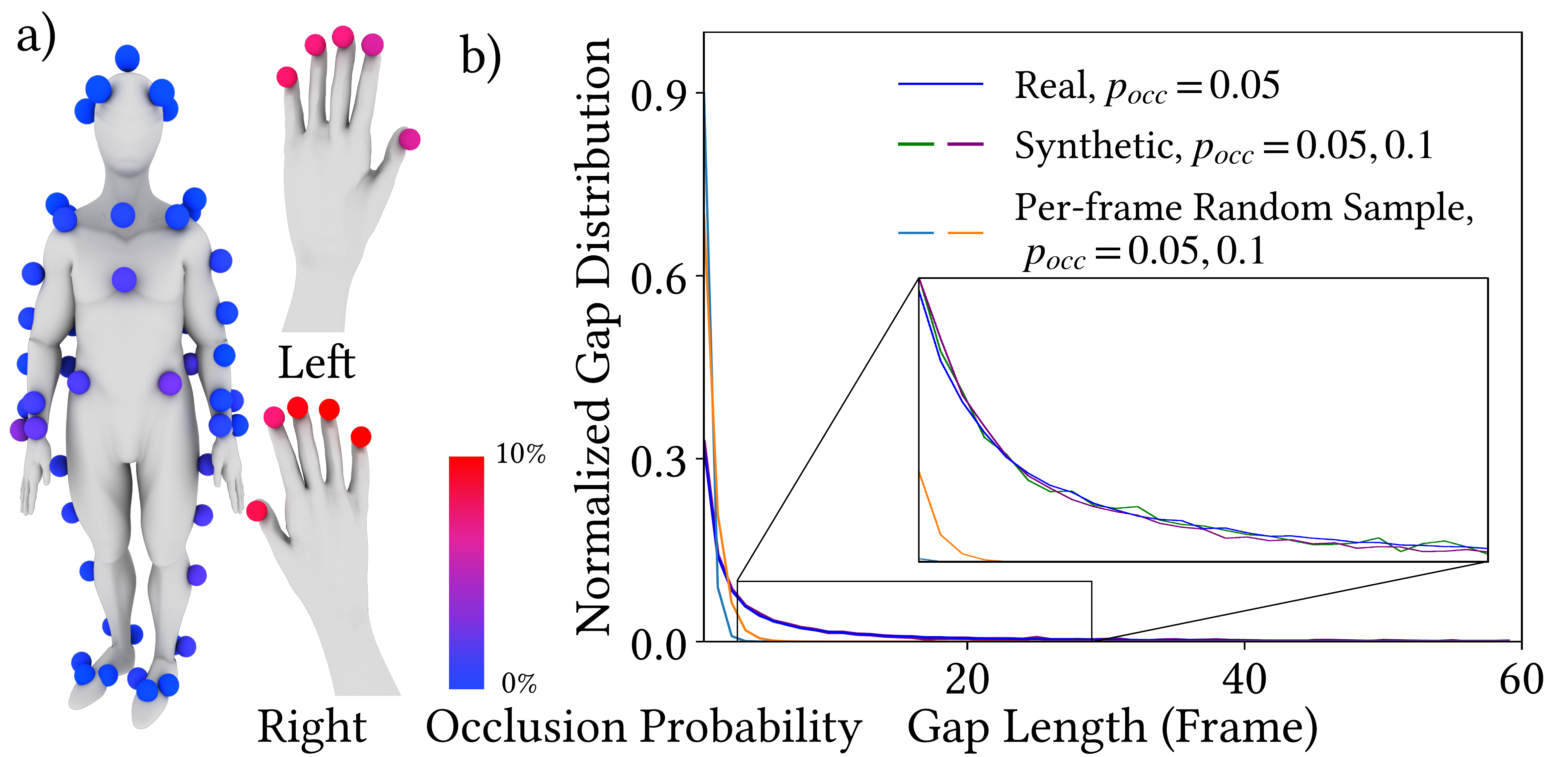}
    \caption{(a) Occlusion probabilities and (b) gap length distribution of the real MoCap data. Markers on the right side of body tend to have higher occlusion probabilities than those on the left side. The distributions of gap length follow a heavy-tailed distribution.}
    \label{gap_distribution}
\end{figure}

One of the most crucial tasks for our networks is to learn to deal with complex occlusions. However, obtaining such noise data is hard due to the high cost of optical MoCap systems and the low frequency of noise occurrences in modern MoCap scenarios. For instance, our real MoCap dataset has an overall occlusion probability of only 5\%. Consequently, there is a limited amount of data available, which makes it easy for the networks to overfit. To address this issue, we draw inspiration from masked autoencoders~\cite{he2022masked} and feed the network with strategically perturbed data. We apply a training regime based on representation learning and data augmentation by simulating shifting and occlusion gaps based on their spatial and temporal patterns. Notably, markers located at different positions on the body exhibit varying occlusion probabilities $p_{occ,i}$ due to differences in motion patterns. For instance, hand markers have higher occlusion probabilities than that of other body parts, as they are most densely placed and the motions of hand are the most complex. Furthermore, because the motion and capture situation is continuous, occlusions usually last for a period of time rather than occurring haphazardly in single frames. The distribution of occlusion gap lengths follows a heavy-tailed distribution that is consistent across different markers. To simplify the process, we use the average gap length distribution of all markers when generating synthetic data. We illustrate the occlusion probabilities $\hat{p}_{occ,i}$ and the occlusion gap length distribution $\hat{g}_l$ collected from real data on Fig. \ref{gap_distribution}. We also illustrate the gap length distribution of our method and per-frame random sample methods~\cite{holden18,MoCapSolver}. Our method generates gaps that are closest to the real distribution, whereas the per-frame random sample method fails to generate gaps longer than 6 frames.

 \CHANGE{Given the overall occlusion probability $p_{occ}$ and total sequence length $L$, we first compute the gap number $N_{l,i}$, which represents the number of gaps with length $l$ for marker $M_i$, using the gap length distribution and marker occlusion probabilities as follows:
}
\begin{equation}
    \begin{split}
        N_{l,i} = {\big\lfloor} \frac{L}{l} \frac{l \hat{g}_l}{\sum (l \hat{g}_l)} p_{occ} \frac{\hat{p}_{occ,i}}{\sum \hat{p}_{occ,i}})  {\big\rfloor} ,
     \end{split}
\end{equation}
where $\frac{l \hat{g}_l}{\sum (l \hat{g}_l)}$ indicates the gap possibility weighted by gap length, $p_{occ} \frac{\hat{p}_{occ,i}}{\sum \hat{p}_{occ,i}}$ represents weighted occlusion probability for marker $M_i$, \CHANGE{and $\lfloor \cdot \rfloor$ represents floor operator.}

Using the calculated occlusion gap numbers $N_{l,i}$, we randomly select segments from longest to shortest without overlapping and set them to occluded. Selecting segments randomly or from shortest to longest may lead to insufficient space for occlusion gaps. After that, we add shifts to markers on randomly sampled frames by applying offsets with the uniform distribution $o \sim U(-\sigma_{shift},\sigma_{shift})$.

\section{Results and Experiments}

We conduct our experiments on two distinct datasets: a real dataset and a synthetic dataset, each featuring characters with different marker configurations and skeleton structures. The real dataset was captured in a game studio, and the synthetic dataset is generated by driving the SMPL+H body~\cite{MANO:SIGGRAPHASIA:2017} using the body motions from the CMU MoCap dataset~\cite{CMU} and the hand motions from the GRAB~\cite{GRAB} dataset. We use 6 neighbor markers with the lowest distance variances for occlusion fixing and 3 for motion solving networks. The supplementary material contains additional information about our dataset, network architectures, hyper-parameters, and other implementation details.

\subsection{Evaluation}

\begin{table*}[htbp]
    \caption{Comparison with other methods and ablations for our method.}
    \scalebox{0.97}{
    \begin{tabular}{cc|cc|cc|cc|cc|cc|cc|cc}
        \multicolumn{2}{c|}{\multirow{2}{*}{}} & \multicolumn{2}{c|}{\makecell{  Optimization-\\ based Methods \\ \scriptsize (Vicon for Real \\ \scriptsize Mosh++ for Synthetic)}} & \multicolumn{2}{c|}{[Holden 2018]} & \multicolumn{2}{c|}{MoCap-Solver} & \multicolumn{2}{c|}{\makecell{Ours \\ \scriptsize (w/o EDM \\ \scriptsize initial value)}} & \multicolumn{2}{c|}{\makecell{Ours \\ \scriptsize (w/o splitting \\ \scriptsize body \& hand)}}      & \multicolumn{2}{c|}{\makecell{Ours \\ \scriptsize (w/o marker \\ \scriptsize convolution)}}      & \multicolumn{2}{c}{Ours}            \\
        
        \multicolumn{2}{c|}{}                  & Body                        & Hand                               & Body                              & Hand                      & Body                     & Hand & Body & Hand & Body & Hand & Body & Hand & Body & Hand                          \\
        \hline
        \multirow{3}{*}{Real}          & JOE (°)                         &   4.21                                 &      1.37                             & 2.54                      & 0.80                     & 1.89 & 0.58 &  1.27    & 0.58  &  \CHANGE{1.31} & \CHANGE{1.17}  & 1.78 & 0.57 &  \textbf{1.22} & \textbf{0.45} \\
                                               & JPE  (cm)                       &  1.75                                  &   0.49                                & 1.02                      & 0.27                     & 0.89 & 0.20 &  0.76    & 0.21     & \CHANGE{0.85} & \CHANGE{0.55} & 0.85 & 0.18 & \textbf{0.61}  & \textbf{0.15} \\
                                               & OMPE (cm)                        &    5.32                                &    2.01                               & 3.62                      & 1.58                     & 3.27 & 1.46 &  3.83  & 1.71     & \CHANGE{2.76} & \CHANGE{1.76}  & 2.60 & 1.15 & \textbf{2.55} & \textbf{1.11} \\
        \hline
        \multirow{3}{*}{Synthetic}     & JOE (°)                        &  5.72                                  & 6.37                                  &        3.53                   & 2.21                         & 2.76    &  1.72    & 2.47  & 1.78    &   \CHANGE{2.25}            &  \CHANGE{3.41}           & 2.65 & 1.70  & \textbf{2.15} & \textbf{1.59}  \\
                                               & JPE  (cm)                       & 2.02                                   &  0.79                                 & 1.37                          &    0.27                      & 1.08     &  0.21    &  1.12   &  0.23  &     \CHANGE{1.17}          &       \CHANGE{0.53}        & 1.09 & 0.20 & \textbf{1.00} & \textbf{0.18} \\
                                               & OMPE  (cm)                       &   7.20                                 &  1.25                                 &    4.58                       &   0.58                       &  4.28    &   0.57   &   5.22   &  0.83    &  \CHANGE{4.02}             & \CHANGE{1.18}       & \textbf{3.66}  & 0.38 & 3.67 & \textbf{0.35}  \\
    \end{tabular}
    }
    \label{comparison_table}
\end{table*}

\begin{figure}[htbp]
    \centering
    \includegraphics[width=\linewidth]{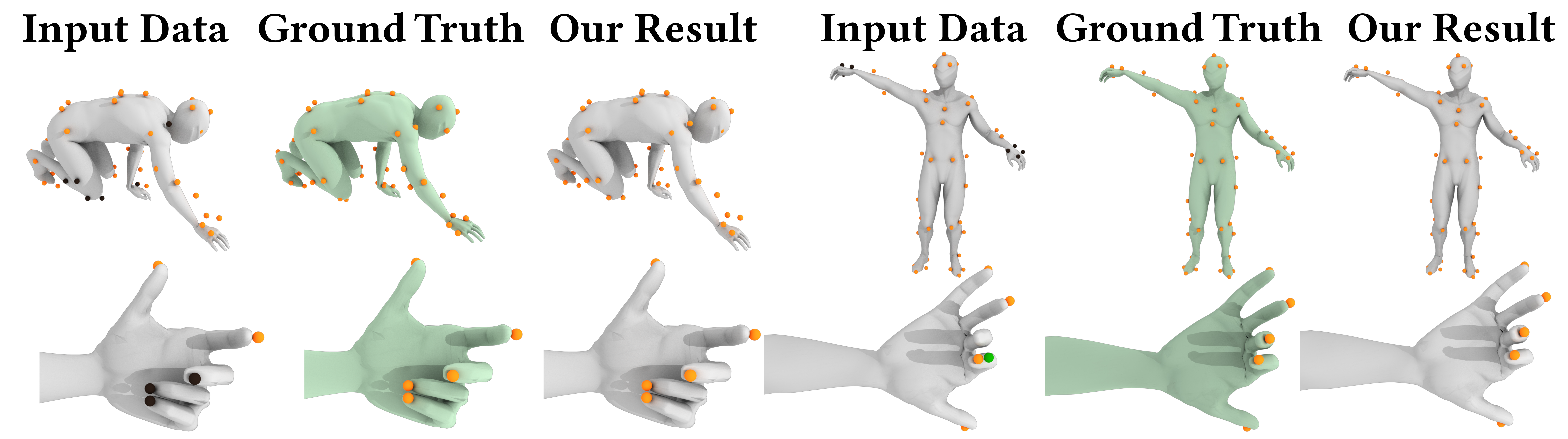}
    \caption{Our method can accurately solve diverse unseen body motions and fine multi-fingered hand motions. Our method is robust to occlusions (black ball) and outliers (green ball).}
    \label{diverse_motions}
\end{figure}

We present the results of our method for unseen motions in Fig. \ref{diverse_motions}. Our method successfully handles diverse motions, including large body movements and fine hand motions. Additionally, our method is robust to marker occlusions and outliers.

\begin{figure}[htbp]
    \centering
    \includegraphics[width=\linewidth]{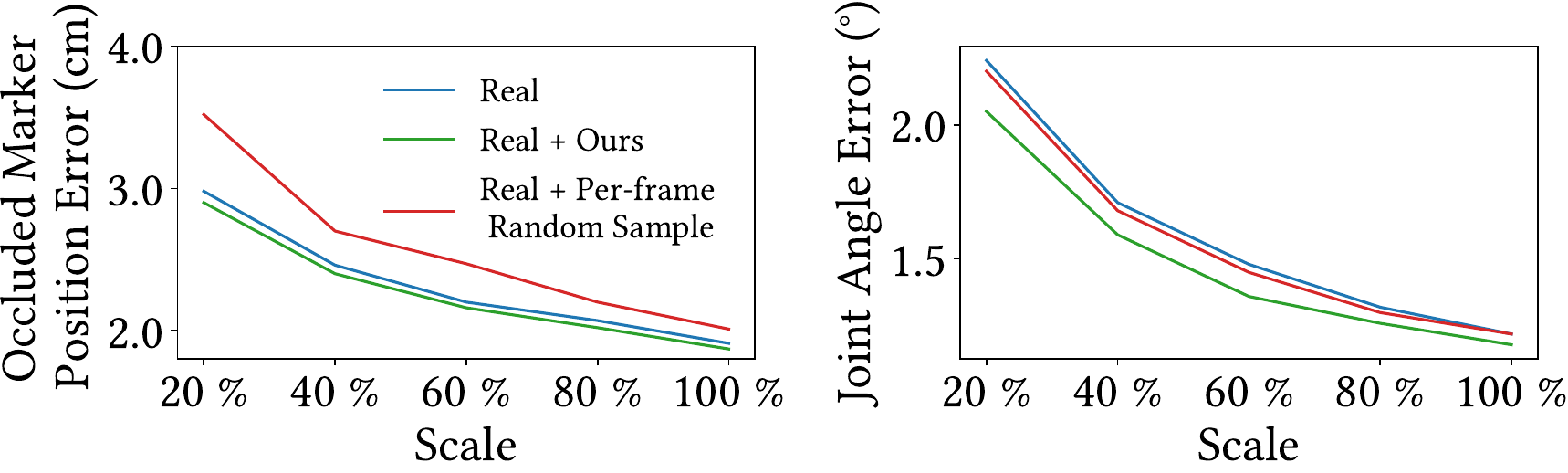}
    \caption{Occluded marker position error and joint angle error vs. training data scale. We also test two data augmentation methods:  \CHANGE{"Real" denotes real dataset without augmentation, whereas "Real + Ours" denotes the same dataset supplemented with data generated using our augmentation method.}}
    \label{data_augmentation}
\end{figure}

Following that, we assess the efficacy of our data augmentation method \CHANGE{on the real dataset}. We randomly sample the training data, selecting 20\%, 40\%, ..., 100\% to train the occlusion fixing and solving networks, and then test them using the same testing set. The quantitative results are shown in Fig. \ref{data_augmentation}. The losses decrease as the data scale increases because the training set contains more diverse data. We also run experiments that augment training data by adding corrupted data from sampled data to the training set. We compare our data augmentation method with the per-frame random sample method used in~\cite{holden18,MoCapSolver}. Our method improves network performance for marker occlusion fixing because it can generate long occlusion gaps with a data distribution similar to real data. The per-frame random sample method degrades the network's performance on real data, as it fails to generate long occlusion gaps.  In terms of solving network, our data augmentation method improves network performance by 10\% when only a small amount of training data is available. The per-frame random sample method, on the other hand, does not significantly improve network performance.

\begin{figure}[htbp]
    \centering
    \includegraphics[width=\linewidth]{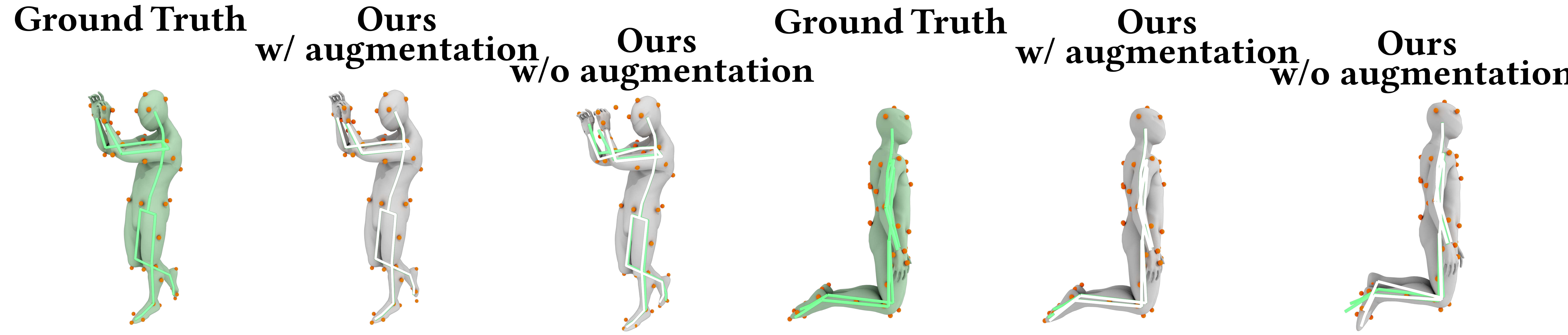}
    \caption{Qualitative comparisons of motions solved using our network trained with 20\% of training set with and without data augmentation.  We render the skeletons and overlay the ground truth's skeleton onto ours.}
    \label{augmentation_quantitative}
\end{figure}

\CHANGE{On the real dataset, we tested different selection criteria and numbers of neighbor markers. The results show that choosing markers with the smallest distance variances as neighbor markers outperforms choosing those with the smallest mean distances. Our approach yields the best results with six markers for occlusion filling and three markers for the network resolution. Please refer to supplementary materia for an in-depth qualitative analysis and discussion.}



We conduct three ablation studies to validate the efficacy of our method's modules. The first study does not use the EDM optimization for the occlusion fixing module, and the second does not separate the body and hand markers and solves their motions in a single network. Finally, the third uses residual modules rather than performing intra-convolutions on the marker graph $\mathcal{G}_M$. The quantitative results are shown on the right side of Table \ref{comparison_table}. For occlusion fixing, the initial value given by EDM improves marker position accuracy by approximately 40\%. As a result of more accurate marker positions, the solving network estimates motions closer to the ground truth with errors reduced by approximately 4\% for the body and 20\% for the hands (where markers have higher occlusion probabilities). Solving whole-body motions with a single network negatively affects the accuracy of both hand and body motions, which justifies the necessity of splitting body and hand markers and solving them using separate networks, which improves hand motion accuracy in terms of angle error by 60\% and body motion accuracy by 5\%. Solving motions by first extracting markers' local features on the marker graph $\mathcal{G}_M$ significantly improves performance of the solving network, reducing errors in terms of angle differences by approximately 30\% for body motions and 15\% for hand motions.

\subsection{Comparison with Prior Methods}

We compare our method with two neural-based approaches: \cite{holden18} and MoCap-Solver~\cite{MoCapSolver} on the two datasets. Additionally, we compare our method with two optimization-based approaches: Vicon~\cite{Vicon} for the real dataset and Mosh++~\cite{AMASS} for the synthetic dataset. The neural-based approaches are trained with the same dataset and evaluated using the same unseen data. To ensure a fair comparison, the body and hand markers are aligned using the same strategy and trained using separate networks for neural approaches. We use three metrics to quantitatively evaluate the reconstructed marker sequence and the solved motions, namely OMPE (Occluded Marker Position Error), JOE (Joint Orient Error), and JPE (Joint Position Error). The first one is the distance between the estimated occluded marker position and the ground truth. The latter two represent different aspects of motion solving accuracy: JOE is the angle difference for joints and JPE is the global joint position difference.

\begin{figure}[htbp]
    \centering
    \includegraphics[width=\linewidth]{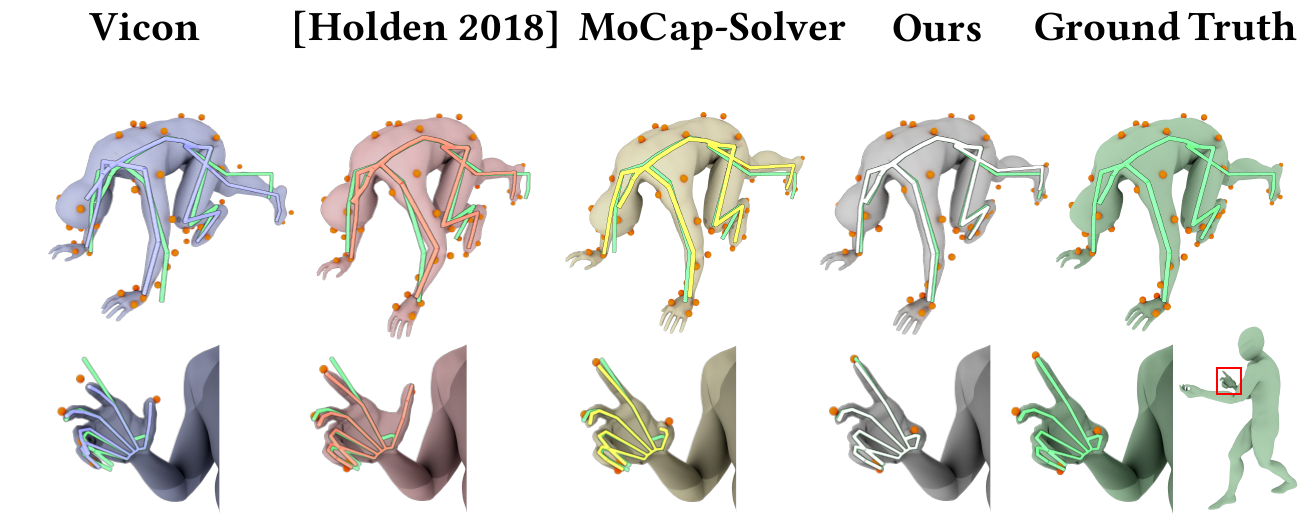}
    \caption{Qualitative comparisons of motions solved using various methods. Top / bottom row: bodies with large motions and hands with fine motions, with the hand marked using red boxes and enlarged for easier comparison. Our method generates motions that are closest to the ground truth.     }
    \label{comparison_little}
\end{figure}

\begin{figure*}[htbp]
    \centering
    \includegraphics[width=\linewidth]{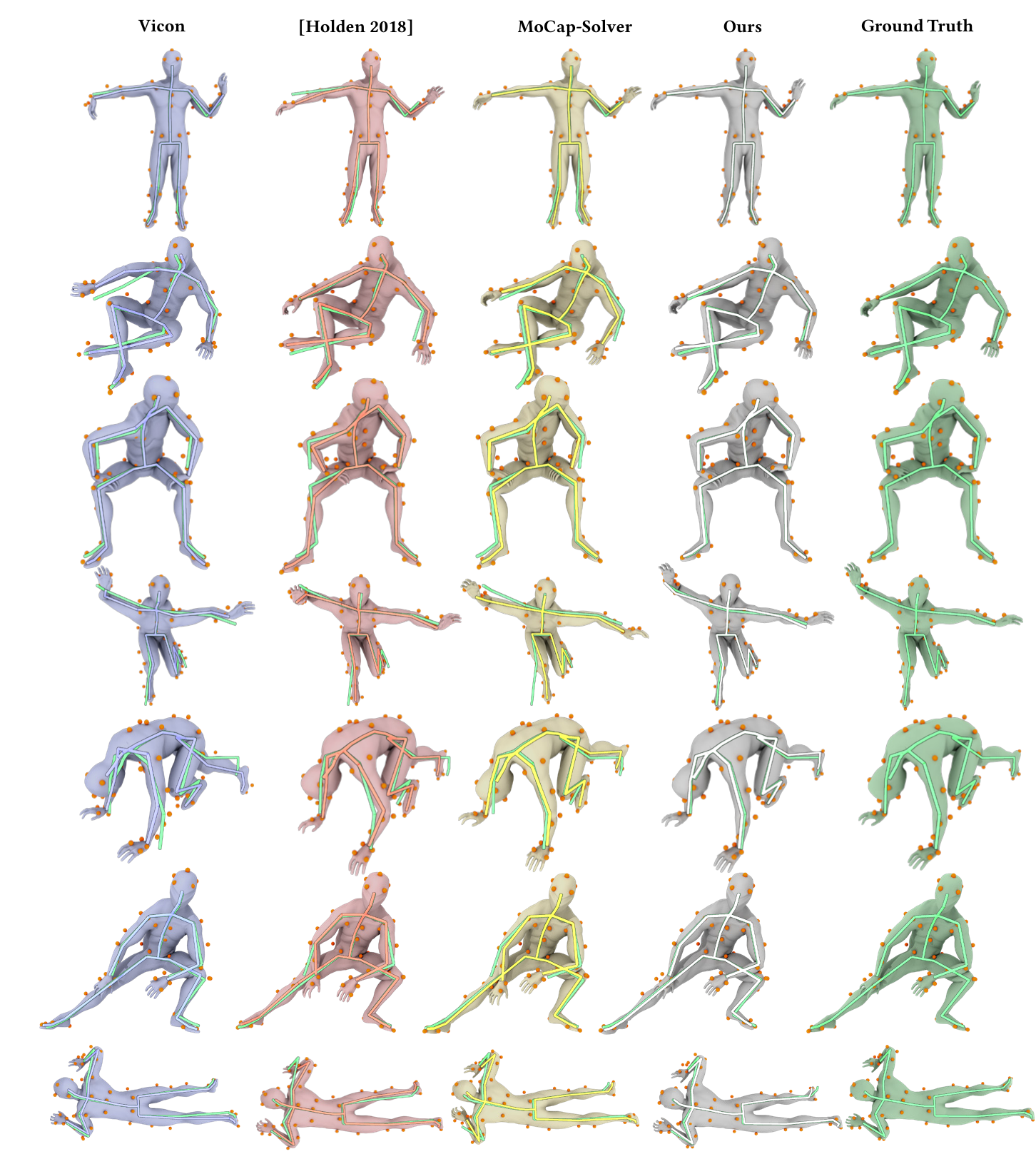}
    \caption{Additional comparison of body motions solved by different methods. To compare with the ground truth, we render the skeletons and overlay the ground truth's skeleton onto those generated by solving methods.}
    \label{additional_results}
\end{figure*}

\begin{figure*}[htbp]
    \centering
    \includegraphics[width=\linewidth]{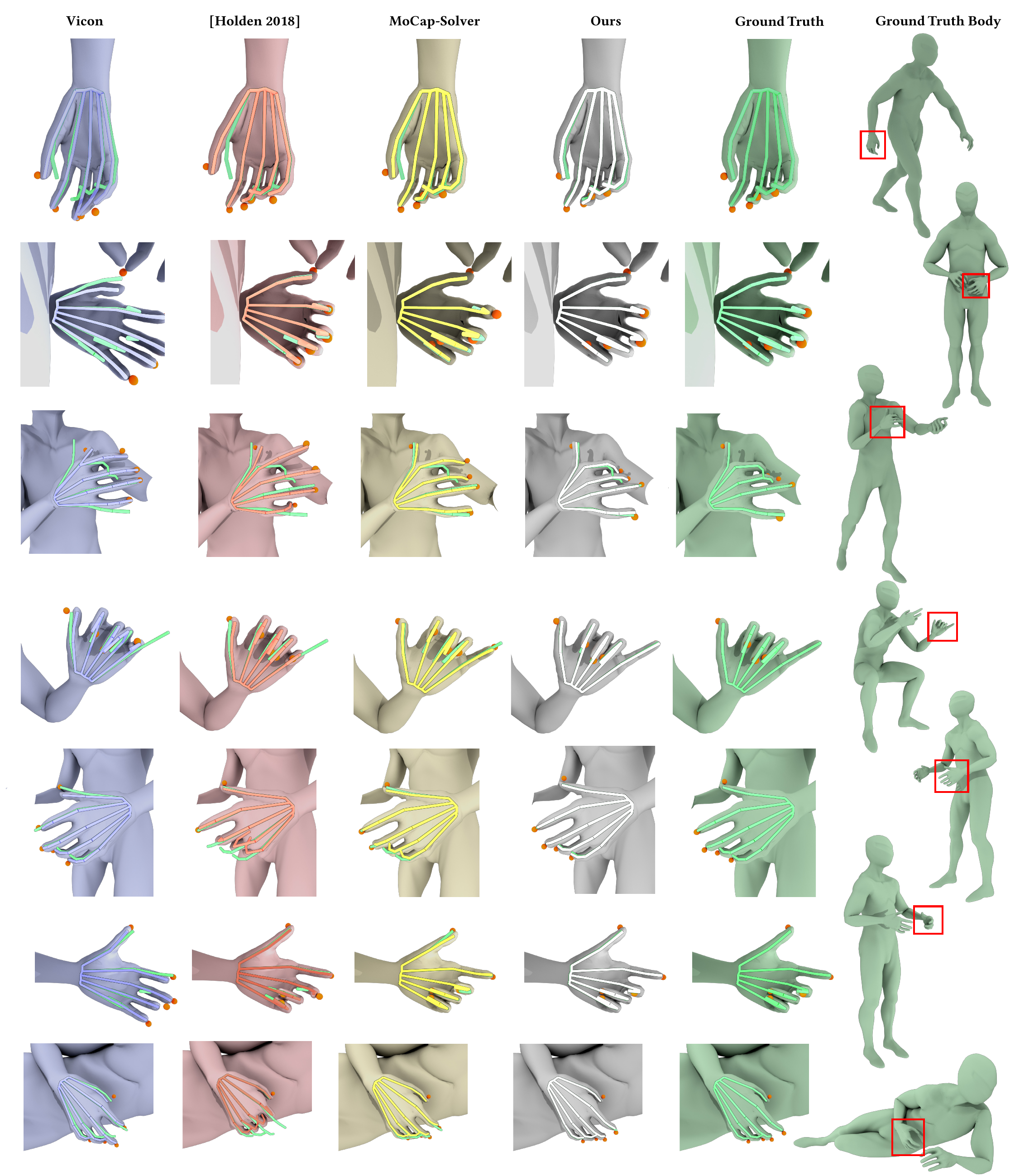}
    \caption{Additional comparison of hand motions solved by different methods. To avoid the interference of body motions, we use the same body motions for different methods. To compare with the ground truth, we render the skeletons and overlay the ground truth's skeleton onto those generated by solving methods.}
    \label{additional_results2}
\end{figure*}

On the left of Table \ref{comparison_table}, we list the quantitative results for comparison methods. Our method outperforms the baseline methods for all metrics. For body motions and markers, JOE and JPE are about 30\% lower, while OMPE is 20\% lower than the best of them. For hand, all metrics are approximately 20\% lower than the state-of-the-art methods. Furthermore, our method achieves lower average errors and generates fewer large errors compared with other methods. We present two detailed comparisons of body and hand motions in Fig. \ref{comparison_little}, where our method tends to generate motions with the highest accuracy. Optimization based methods~\cite{Vicon,AMASS} lack the learned prior knowledge for the occluded markers and motions, which fails to accurately solve motions when large numbers of markers are occluded. For two reasons, our method outperforms other neural-based methods~\cite{holden18,MoCapSolver}. First, our method better reconstructs the positions of the occluded markers, providing a better input to the solving network. Second, our solving network extracts local features of each marker and joint, which aids in generating motions that are close to the ground truth. We present additional qualitative comparisons of large body motions on Fig. \ref{additional_results} and fine hand motions on Fig. \ref{additional_results2}.

\section{Conclusion, Limitations, and Future Work}

We present a learning-based method that leverages locality to clean optical MoCap data and solve body and hand motions. We have evaluated it on diverse and complex benchmarks and highlighted improved performance over prior baseline methods. Compared with previous methods, our method better recovers the occluded markers' positions and more accurately solves the body and hand motions. Additionally, our method generates synthetic MoCap data with distributions close to real data.

Our method has limitations. First, our data augmentation method solely considers the spatiotemporal distribution of occlusions. Incorporating other occlusion and noise patterns, such as simultaneous occlusion, varying shifting intensities of markers on different parts of body, and mislabeling between close markers, may yield a distribution more closely resembling real data. Second, our method is unable to detect subtle marker swaps, such as those between the index and middle fingers. Detecting and correcting such exchanges is a worthwhile future endeavor.

\begin{acks}
Xiaogang Jin was supported by Key R\&D Program of Zhejiang (No. 2023C01047) and the National Natural Science Foundation of China (Grant Nos. 62036010, 61972344).  We thank Tao Feng, Yuchen Liu, Shaohua Niu, Xuetong Meng, Qishen Liu, Cheng Ge, and other colleagues in Tencent Games Digital Content Technology Center for preparing the training data, discussing the result and rendering the video.


\end{acks}

\bibliographystyle{ACM-Reference-Format}
\bibliography{refrences}

\end{document}